\RequirePackage{fix-cm}

\documentclass[onecolumn]{svjour3}

\usepackage[utf8]{inputenc}
\usepackage{graphicx}
\usepackage{amsmath,mathrsfs}
\usepackage{amssymb}
\usepackage{epstopdf}
\usepackage{braket}
\usepackage{bbm}
\usepackage{xcolor}

\newcommand{\be}{\begin{equation}}
\newcommand{\ee}{\end{equation}}
\newcommand{\beq}{\begin{eqnarray}}
\newcommand{\eeq}{\end{eqnarray}}
\newcommand{\Tr}{\mathrm{Tr}}

\journalname{Foundations of Physics}

\begin{document}

\title{Hardy's paradox as a demonstration of quantum irrealism}


\author{Nicholas G. Engelbert \and Renato M. Angelo}

\authorrunning{Engelbert and Angelo}

\institute{N. G. Engelbert \at
          Department of Physics, Federal University of Paran\'a, P.O. Box 19044, 81531-980 Curitiba, Paraná, Brazil.
          \and
          R. M. Angelo \at
          Department of Physics, Federal University of Paran\'a, P.O. Box 19044, 81531-980 Curitiba, Paran\'a, Brazil. \\
          Tel.: +55-41-33613092 \\
          Fax: +55-41-33613418\\
          \email{renato@fisica.ufpr.br}
}

\date{Received: date / Accepted: date}

\maketitle

\begin{abstract}
Hardy's paradox was originally presented as a demonstration, without inequalities, of the incompatibility between quantum mechanics and the hypothesis of local causality. Equipped with newly developed tools that allow for a quantitative assessment of realism, here we revisit Hardy's paradox and argue that nonlocal causality is not mandatory for its solution; quantum irrealism suffices.
\keywords{Hardy's paradox \and Realism \and nonlocality}
\end{abstract}

\section{Introduction}
\label{intro}

Motivated by the Einstein, Podolsky, and Rosen (EPR) conclusion~\cite{epr} that quantum mechanics is an incomplete theory, Bell showed that no theory intended to restore causality and locality can be consistent with the statistical results of quantum mechanics~\cite{bell}. The hypothesis of {\it local causality} considered by Bell can be written as 
\be 
p(a,b|A,B)=\sum_\lambda p_\lambda\,p(a|A,\lambda)\,p(b|B,\lambda),
\label{lc}
\ee 
where $p(a,b|A,B)$ is an experimentally accessible joint probability distribution of finding outcomes $a$ and $b$ for measurements of observables $A$ and $B$ in the systems $\mathcal{A}$ and $\mathcal{B}$, respectively, $\lambda$ is a hidden variable satisfying $\sum_\lambda p_\lambda=1$, and $p(a|A,\lambda)$ and $p(b|B,\lambda)$ are marginal probability distributions. From the above hypothesis, inequalities can be derived---such as the Clauser-Horne-Shimony-Holt (CHSH) inequality~\cite{chsh}---and tested in laboratory. Several loophole-free experiments performed in the recent past \cite{giustina,hensen_2015,shalm2015,hensen_2016,rauch2018,li2018} have convincingly shown that nature is, in agreement with quantum mechanical predictions, incompatible with the local causality hypothesis. In this work, we employ the term {\it nonlocal causality} to make the antithesis with local causality, whose inadequacy is by now a fact. With that we intend to stress the interpretation according to which there must be in play some ``spooky action at a distance'', that is, cause and effect are instantaneously connected via some nonlocal interaction.

\begin{figure}[htb]
\centerline{\includegraphics[width=0.6\textwidth]{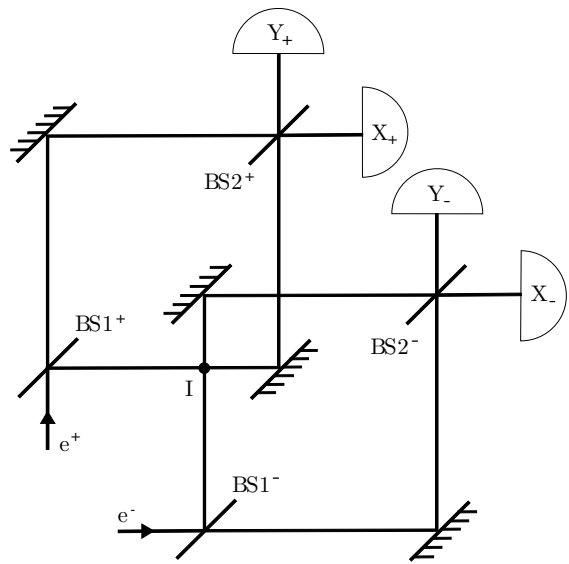}}
\caption{\small Depiction of Hardy's experiment. A Mach-Zehnder interferometer MZ$_+$ (MZ$_-$) is projected to ensure that incoming positrons (electrons) never reach the dark detector X$_+$ (Y$_-$). When MZ$_+$ and MZ$_-$ are put together, positron-electron annihilation is certain to occur if the particles meet at point I. Hardy's paradox occurs when a dark detector clicks, for in this case the particles took nonoverlapping paths and interacted at a distance, thus violating the classical premise of local causality.}
\label{fig1}
\end{figure}

In 1992, Hardy introduced a gedankenexperiment (depicted in Fig.~\ref{fig1}) to demonstrate Bell's theorem without using inequalities~\cite{hardy}. In the setting, two Mach-Zehnder interferometers denoted MZ$_\pm$ are arranged so that two arms overlap. For the upper (lower) interferometer, MZ$_+$ (MZ$_-$), the incoming particle is a positron (electron). If the particles meet at point I, which happens with probability $\tfrac{1}{4}$, then positron-electron annihilation occurs, $\gamma$-radiation is generated, and no detector clicks. The interferometer MZ$_+$ (MZ$_-$) is calibrated in a way such that, when the other interferometer is far apart, then detector X$_+$ (Y$_-$) works as a ``dark detector'', that is, it never clicks. The paradoxical instance then emerges when at least one dark detector clicks: since annihilation did not occur, the particles could not both have travelled the overlapping region, but they must have ``felt'' one another at a distance, otherwise the dark detectors could not have clicked. This rationale, firmly based on the premise of well-defined trajectories (realism), demonstrates nonlocal causality.

Interpretative matters aside, a purely statistical analysis of all possible paths predicts that, with probability $\tfrac{3}{16}$, at least one dark detector clicks and, with $\tfrac{1}{16}$, both click. Most importantly, numerous experimental works verified these predictions for discharged quantum systems (photons)~\cite{exp0,exp1,exp2,exp3,exp4,exp5,exp6,exp7,exp8,exp9}. While Hardy's setting was not the first demonstration of Bell's theorem without inequalities \cite{ghsz}, his was the first designated for a two-part system with two dichotomic measurements (CHSH scenario). Hardy showed that his approach applies for a wide class of entangled states in the two-part case~\cite{hardy2}, and since then other developments were made in generalizing the paradox for the $n$-part case~\cite{cereceda_2004,jiang} and  higher dimensional local systems~\cite{chen}. Also, analogous paradoxes for the CHSH scenario were constructed in Ref.~\cite{fritz}, and a unification of the CHSH inequality with Hardy's experiment was constructed in Ref.~\cite{hardy_chsh} through the framework of nonlocal games.

Hardy's experiment challenges the usual notion of what is considered a ``local interaction''. If one of the ``branches'' of a superposition state interacts with another system, under what assumptions may that interaction be considered local? If that ``branch'' is then led into interference, under what assumptions may we even assert that the interaction did take place, without resorting to some sort of counterfactual reasoning? Aharonov {\it et al.} \cite{aharonov} revisited the paradox utilizing the framework of weak measurements to analyze the state of the system before reaching the final beam-splitters, while arguing against the claims of counterfactuality that are commonly used to dismiss the paradox. Here, we concur with the view of these authors that analysis of the quantum states inside of the interferometer may bring interesting insights into the foundational issues involved in the paradox, even though no measurement is performed at that stage.

In this work, we defend that the results underlying Hardy's experiment can be interpreted in totally the opposite way. Instead of conceiving that the particles are always traveling well-determined paths (realistic trajectories subjectively ignored by the observer) and can interact at a distance (nonlocal causality), we abandon the notion of realism and admit that interactions are always local\footnote{It is worth noticing that Hardy himself suggested in his 1992 paper~\cite{hardy} that abandoning realism would be another way out of the paradox. This suggestion was supported by the demonstration, given in the very same work, that realistic theories cannot be Lorentz invariant. It seems to us, however, that, despite the relevance of such argument, viewing Hardy's paradox as a demonstration of nonlocality has been the prevalent position among physicists.}. For this purpose, we adopt a notion of realism recently introduced in the literature. This new (nonsignalling) framework does not amount to banishing nonlocality from nature, but the emerging nonlocal aspects are, much like Bell nonlocality, in perfect conformity with special relativity.

\section{Elements of reality}
\label{reality}

In their seminal 1935 paper~\cite{epr}, EPR put forth their criterion of physical reality. Roughly speaking, the criterion states that whenever a complete theory ensures full predictability for the value of an observable prior to any disturbance of the system, then this observable is an element of reality. Together with the assumption of locality, this criterion would lead EPR to claim that for the singlet state, $\ket{\mathfrak{s}}=\tfrac{1}{\sqrt{2}}\left(\ket{+-}-\ket{-+} \right)$, the spins of the particles, in every measurement direction, are all elements of reality. Since then, many alternative criteria have been suggested~\cite{bohr,ruark,redhead,bruk_zei1,bruk_zei2,zeilinger_1999,qbism,krizek}. 
Recently, Bilobran and Angelo (BA) introduced an operational criterion of realism \cite{bil} based on the single premise that after an observable is measured, then it becomes an element of reality. The construction goes as follows. Consider that a task is given to an experimentalist to find out, via ideal state tomography, the multipartite state $\rho\in\mathcal{B(H_A\otimes H_B)}$ (with $\mathcal{H_B}=\bigotimes_{i=2}^N\mathcal{H}_i$) prepared by some source. The source prepares infinitely many copies of $\rho$ and the experimentalist is allowed to make as many measurements as needed. Now, in every run of the experiment, a secret agent intercepts the state right after its preparation and always measures the same discrete-spectrum observable $A=\sum_aaA_a$, where $A_a=\ket{a}\bra{a}\in\mathcal{B(H_A)}$, leaving the other parts of the system untouched. After the measurement is conducted and the outcome $a$ is obtained, the preparation $\rho$ collapses to $A_a\otimes\rho_{\mathcal{B}|a}$, where $\rho_{\mathcal{B}|a}=\Tr_\mathcal{A}[(A_a\otimes\mathbbm{1}_\mathcal{B})\rho (A_a\otimes\mathbbm{1}_\mathcal{B})]/p_a$ and $p_a=\Tr[(A_a\otimes\mathbbm{1}_\mathcal{B})\rho (A_a\otimes\mathbbm{1}_\mathcal{B})]$. The measurement outcomes are never revealed to the experimentalist. After the completion of the protocol, the experimentalist obtains the result
\be 
\sum_ap_aA_a\otimes\rho_{\mathcal{B}|a}=\sum_a(A_a\otimes\mathbbm{1}_\mathcal{B})\rho(A_a\otimes\mathbbm{1}_\mathcal{B})=:\Phi_A(\rho),
\label{Phi}
\ee 
which corresponds to the collection of collapsed states pondered by their respective probabilities. The object $\Phi_A$ denotes a completely positive trace-preserving unital map that formally refers to the above protocol of unrevealed measurements. BA then use their premise to ascribe to $\Phi_A(\rho)$ the connotation of {\it $A$-reality state}, that is, a state for which $A$ is an element of reality. Remarkably, one verifies that $\Phi_A\Phi_A(\rho)=\Phi_A(\rho)$, meaning that further measurements of $A$ cannot disturb a state for which $A$ is already an element of reality. From this observation, BA introduce their criterion of realism:
\be 
\Phi_A(\rho)=\rho\qquad\qquad \text{(BA's criterion of realism)}.
\label{BAcriterion}
\ee 
Clearly, this criterion makes reference to the realism associated with a particular observable. Two cases are particularly noteworthy, namely, $\Phi_A(A_a)=A_a$ and $\Phi_A(\rho_\mathfrak{s})\neq\rho_\mathfrak{s}$, for $\rho_\mathfrak{s}=\ket{\mathfrak{s}}\bra{\mathfrak{s}}$ and any spin-$\tfrac{1}{2}$ operator $A=\vec{u}\cdot\vec{\sigma}$, with $||\vec{u}||=1$. For the former, EPR's criterion agrees with that of BA, since given the eigenstate $\ket{a}$ of $A$ we can certainly predict the outcome for an eventual measurement of this observable. For the latter, however, the two criteria are in dramatic disagreement: while EPR's criterion claims full realism (for all spin-$\tfrac{1}{2}$ observables), BA's predicts just the antithesis, that is, full {\it irrealism}. 

The framework introduced by BA included the measure called {\it irreality},
\be
\mathfrak{I}_A(\rho):=S(\Phi_A(\rho))-S(\rho),
\label{frakI}
\ee
which quantifies, via the ``metric'' induced by the von Neumann entropy $S$, by how much $\rho$ is far from an $A$-reality state $\Phi_A(\rho)$ for a given context $\{A,\rho\}$. This measure is always nonnegative and vanishes if and only if the realism criterion \eqref{BAcriterion} is satisfied. As shown in Ref.~\cite{bil}, this measure can be readily decomposed as $\mathfrak{I}_A(\rho)=\mathfrak{I}_A(\rho_\mathcal{A})+D_A(\rho)$. The first parcel refers to the {\it local irreality}, that is, the violation degree of BA's criterion given that only the reduced state $\rho_\mathcal{A}$ is accessed. It has been shown that this quantity can be related to quantum coherence in the $A$ basis~\cite{baum} and  waviness~\cite{angelo2015}. The second parcel is the basis-dependent quantum discord (a measure of quantum correlations associated with the observable $A$).  Such decomposition implies that $\mathfrak{I}_A(\rho)-\mathfrak{I}_A(\rho_\mathcal{A})\geqslant \mathfrak{D}_\mathcal{A}(\rho)$, where $\mathfrak{D}_\mathcal{A}(\rho):=\min_AD_A(\rho)$ is the (one-way) quantum discord~\cite{ollivier,henderson}. This inequality reveals that quantum correlations forbid irreality to be equivalent to local irreality, meaning that the $A$-realism cannot be devised without reference to other parts of the system. 

Further developments have recently been put forward with regard to the BA approach to realism. In Ref.~\cite{dieguez2018} a complementarity relation was derived between irreality and quantum information, with experimental tests being implemented via a photonic platform~\cite{mancino2018}. In Ref.~\cite{freire2019} the irreality formalism was extended to continuous variables and the ``uncertainty relation'' 
\be 
\mathfrak{I}_A(\rho)+\mathfrak{I}_{A'}(\rho)\geqslant S\left(\rho||\tfrac{\mathbbm{1}_\mathcal{A}}{d_\mathcal{A}}\otimes\rho_\mathcal{B}\right)
\ee
derived, where $S(\varrho||\sigma)$ is the relative entropy of $\varrho$ and $\sigma$, $\rho_\mathcal{B}=\Tr_\mathcal{A}(\rho)$ is reduced state of part $\mathcal{B}$, and $\{A,A'\}\in\mathcal{B(H_A)}$ are arbitrary observables. The inequality shows that, except when $\rho=\tfrac{\mathbbm{1}_\mathcal{A}}{d_\mathcal{A}}\otimes\rho_\mathcal{B}$, full realism is prevented. One of the most intriguing consequences of BA's approach is the so-called {\it contextual realism-based nonlocality},
\be
\mathcal{N}_{AB}(\rho):=\mathfrak{I}_A(\rho)-\mathfrak{I}_A(\Phi_B(\rho)).
\label{etaAB}
\ee
Being always nonnegative and vanishing for fully uncorrelated states ($\rho=\rho_\mathcal{A}\otimes\rho_\mathcal{B}$) and reality states ($\rho=\Phi_A(\rho)$ or $\rho=\Phi_B(\rho)$), this measure quantifies alterations of the $A$-irreality in the site $\mathcal{A}$ induced by unrevealed measurements of $B$ performed in a remote site $\mathcal{B}$ for a given preparation $\rho$. Many aspects are now well established for the above concept. In Ref.~\cite{valber}, the measure $\mathcal{N}(\rho)=\max_{A,B}\mathcal{N}_{AB}(\rho)$ was introduced as a genuine nonanomalous quantifier of realism-based nonlocality and shown to be fundamentally different from Bell nonlocality. In addition, it has been shown that $\mathcal{N}(\rho)$ is rather resilient to local disturbances, occupies a peculiar position within a given hierarchy of nonclassicality quantifiers~\cite{valber2}, is the only nonclassical aspect that survives in the asymptotic dynamics of two noninteracting quantum walkers~\cite{orthey}, admits a formulation for tripartite states, and is monogamous in some scenarios~\cite{fucci}.

For the purposes of this work, two remarks are now opportune. First, for a reality state like $\varrho=\sum_\lambda p_\lambda\,A_\lambda'\otimes B_\lambda'=\Phi_{A'}(\varrho)=\Phi_{B'}(\varrho)$ one may prove~\cite{valber2} that $\mathcal{N}_{AB}(\varrho)=H(\{p_\lambda\})$, with $H$ the Shannon entropy, for observables $\{A,B\}$ maximally incompatible with $\{A',B'\}$. This means that realism for a context $\{A',B'\}$ does not prevent realism-based nonlocality for other contexts. Furthermore, for realism-based nonlocality to manifest itself in a context $\{A,B\}$, quantum irrealism is necessary for both $A$ and $B$, that is, $\mathfrak{I}_A(\rho)>0$ and $\mathfrak{I}_B(\rho)>0$. Second, the local causality hypothesis \eqref{lc} has no clear link with realism~\cite{gisin}. This can be formally demonstrated within the BA framework. Let us confine the hypothesis \eqref{lc} to the quantum mechanical realm, where the marginal probability distributions are written as $p(a|A,\lambda)=\Tr(A_a\rho^\mathcal{A}_{\lambda})$ and $p(b|B,\lambda)=\Tr(B_b\rho^\mathcal{B}_{\lambda})$. In this case one can write $p(a,b|A,B)=\Tr(A_a\otimes B_b\rho_\text{sep})$, where $\rho_\text{sep}=\sum_\lambda p_\lambda\rho_\lambda^\mathcal{A}\otimes\rho_\lambda^\mathcal{B}$ (a separable state). This unentangled state clearly satisfies local causality, but cannot be claimed to be a reality state since, in general, $\Phi_{A(B)}(\rho_\text{sep})\neq\rho_\text{sep}$.

\section{Reassessing Hardy's paradox}
\label{paradox}

The usual description of Hardy's experiment consists of considering, as pertinent degrees of freedom, the orthogonal paths taken by the electron and the positron, so that a two-dimensional Hilbert space is ascribed to each one of these systems. In our approach, the basis states $\ket{x}_\pm$ and $\ket{y}_\pm$ make reference to the directions the particles travel in their respective interferometers (see Fig.~\ref{fig1}), where the code $\{x,y\}=\{\text{horizontal},\text{vertical}\}$ is reserved for the path direction, and $\{+,-\}=\{\text{positron},\text{electron}\}$ for the matter system. The first conceptual difference with respect to the original treatment of the problem is the inclusion of a state space for the photons generated via annihilation. To this end, we consider a two-dimensional description defined by the states $\ket{0}_\gamma$ (no photon) and $\ket{2}_\gamma$ (two photons), corresponding to the pre- and post-annihilation scenarios, respectively, and spanning the Hilbert space of the photon, $\mathcal{H}_\gamma$. Naturally, our description of the matter system must now be reviewed in order to account for our newly included degree of freedom. We introduce the states $\ket{0}_\pm$ to describe the ``absence'' of matter, so that the particles are now described by three-dimensional Hilbert spaces $\mathcal{H}_\pm$. Therefore, our model deals with a space state $\mathcal{H=H_+\otimes H_-\otimes H_\gamma}$ with dimension $\dim{\mathcal{H}}=18$. It is opportune to point out that the notation here adopted encodes at once spatial and energetic degrees of freedom, that is, the vector $\ket{x}_+$, for instance, refers to a system with rest energy $mc^2$ and elementary charge $e>0$ traveling in some horizontal arm of the interferometer MZ$_+$, whereas $\ket{0}_+$ denotes that the positron has gone, so that its rest energy is no longer available, and no particular traveling direction makes sense anymore. Hence, $\langle 0|x\rangle_+=0$.

With these conventions, the local action of the beam-splitters is formally described by the mapping
\be\begin{array}{lcl}
  \ket{x}_{\pm} &\mapsto& \displaystyle \frac{\ket{x}_{\pm}+i\ket{y}_{\pm}}{\sqrt{2}}, \\
  \ket{y}_{\pm} &\mapsto& \displaystyle\frac{\ket{y}_{\pm}+i\ket{x}_{\pm}}{\sqrt{2}},
\end{array}\label{b-s}\ee
while the local action of the mirrors is given by
\be\begin{array}{lcl}
  \ket{x}_{\pm} &\mapsto& i\ket{y}_{\pm}, \\
  \ket{y}_{\pm} &\mapsto& i\ket{x}_{\pm}.
\end{array}\label{mirror}\ee
The tunings of the interferometers are such that the output directions are the same as the input ones, that is, if the electron, for instance, is prepared in the state $\ket{x}_-$ and no positron is present, then it always reaches the X$_-$ detector, while the Y$_-$ detector never clicks. The same reasoning applies to the positron, which is prepared in the state $\ket{y}_+$. The annihilation process is here prescribed in terms of a fundamentally {\it local} interaction. Mathematically, this is implemented with the mapping $\ket{x}_+\ket{y}_-\ket{0}_\gamma \mapsto \ket{0}_+\ket{0}_-\ket{2}_\gamma$, which is to be applied only when the particles meet each other at the point I. For the sake of generality, however, we model the interaction as
\be
\ket{x}_+\ket{y}_-\ket{0}_\gamma \quad\mapsto\quad \alpha\ket{x}_+\ket{y}_-\ket{0}_\gamma + \beta\ket{0}_+\ket{0}_-\ket{2}_\gamma\qquad \text{(at point I),}
\label{interaction}
\ee
where $\alpha\equiv\sqrt{1-p}\in\mathbb{R}$, $\beta\equiv \sqrt{p}\,\mathrm{e}^{i\varphi}$, $\varphi$ is a generic phase, and $p\in[0,1]$. Also, it is assumed that nothing happens when the particles do not pass through the overlapping region\footnote{Of course, we are artificially ``turning off'' both the Coulomb and the gravitational interactions between the positron and the electron. This is not a big deal for actual experiments, which usually employ photonic platforms.}. In this picture, annihilation occurs with probability $p$, while there is a probability $1-p$ of nothing happening. The amplitudes $\alpha$ and $\beta$ may be thought of as being related, for example, to the cross section of the scattering process at hand. Attention should be drawn to the limiting cases $p=0$ and $p=1$: the former corresponds to no interaction at all (as in a setting in which the interferometers are set far apart from each other), while the latter restores Hardy's original picture, in which annihilation is certain when the particles meet. With this model, we allow for the description of more general scenarios where the radiation can get entangled with the matter. It is noteworthy that no hypothesis whatsoever is invoked regarding ``actions at a distance'', that is, we are definitely excluding from our approach, by construction, nonlocal causality.

We are now ready to analyze aspects of realism and realism-based nonlocality in the whole dynamics of the particles through the interferometers. To this end, we have to choose the observable whose degree of realism we want to diagnose. Consider the positron basis $\{\ket{x}_+,\ket{y}_+,\ket{0}_+\}$. Since the first two vectors do not distinguish between energy states, we naturally adopt {\it path}, henceforth denoted $\mathcal{P}_+$ ($\mathcal{P}_-$ for the electron), as our figure of merit. This quantity refers to the arm (or the direction) which the system is traveling and, therefore, provides information about spatial localization. With that, we now divide the experiment dynamics into four distinct stages and compute, for each stage $k$, the global state of the system, $\ket{\Psi_k}\in\mathcal{H}$, the positron-electron reduced state, $\rho_k=\Tr_\gamma(\ket{\Psi_k}\bra{\Psi_k})$, the irrealities, $\mathfrak{I}_{\mathcal{P}_\pm}(\rho_k)$, the local irrealities, $\mathfrak{I}_{\mathcal{P}_\pm}(\rho_k^\pm)$, with $\rho_k^\pm=\Tr_\mp(\rho_k)$, and the contextual realism-based nonlocality $\mathcal{N}_{\mathcal{P_+P_-}}(\rho_k)$.

\subsection*{\it {\bf Stage 1.} Region before the first set of beam-splitters.}
The initial state of the whole system reads 
\be 
\ket{\Psi_1}=\ket{y}_+\ket{x}_-\ket{0}_\gamma, 
\label{Psi1}
\ee 
while the matter reduced state is $\rho_1=\ket{y}\bra{y}_+\otimes\ket{x}\bra{x}_-$. This state is pure and separable, and both positron and electron are found in eigenstates of $\mathcal{P}_\pm$. Therefore, the state satisfies the criterion given in Eq. \eqref{BAcriterion} and the paths are completely real, that is, $\mathfrak{I}_\mathcal{P_\pm}(\rho_1)=\mathfrak{I}_\mathcal{P_\pm}(\rho_1^\pm)=0$. From the separability of the state, it follows that $\mathcal{N_{P_+P_-}}(\rho_1)=0$. At this stage, therefore, a fully classical worldview is admissible: there is neither irrealism nor nonlocality.

\subsection*{\it {\bf Stage 2.} Region after the initial beam-splitters and before point I.}
After the system passes through the first beam-splitters, the state evolves to 
\be 
\ket{\Psi_2}=\left(\tfrac{\ket{y}_++i\ket{x}_+}{\sqrt{2}}\right)\left(\tfrac{\ket{x}_-+i\ket{y}_-}{\sqrt{2}}\right)\ket{0}_\gamma\equiv\ket{\phi_2^+}\ket{\phi_2^-}\ket{0}_\gamma, 
\ee 
where $\ket{\phi_2^+}\ket{\phi_2^-}$ denotes the (separable) positron-electron state that would emerge if the interferometers were far apart from each other. Since $\rho_2$ is separable, one has $\mathcal{N_{P_+P_-}}(\rho_2)=0$. The action of the beam-splitters adds coherence (wavelike behaviour) into the system, so it is expected that the paths get more \textit{indefinite} at this stage. In fact, we have $\mathfrak{I}_\mathcal{P_\pm}(\rho_2) = \mathfrak{I}_\mathcal{P_\pm}(\rho_2^{\pm}) = \ln 2$. The equivalence between irreality and local irreality reflects the fact that the positron and the electron share no correlations, and the value $\ln 2$ tells us that the paths are maximally unreal. Therefore, according to the present framework, here we have to abandon realism.

\subsection*{\it {\bf Stage 3.} Region after point I and before the mirrors.}
\begin{figure}[htb]
\centering
\includegraphics[scale=0.45]{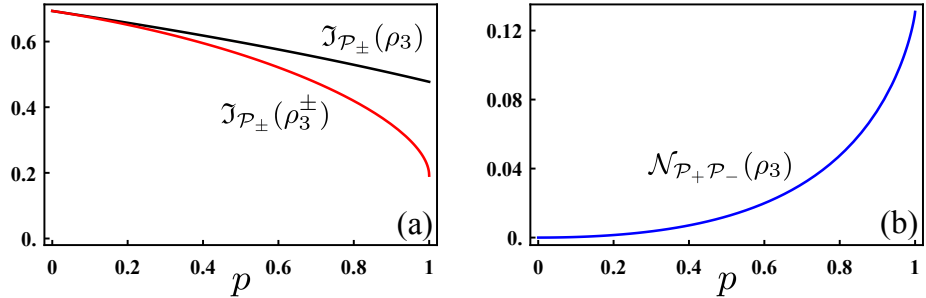}
\caption{\small (a)~Path irrealities $\mathfrak{I}_\mathcal{P_\pm}(\rho_3)$ (upper black line) and path local irrealities $\mathfrak{I}_\mathcal{P_\pm}(\rho_3^\pm)$ (lower red line) at stage 3 of Hardy's experiment as a function of the annihilation probability $p$. (b)~Contextual realism-based nonlocality $\mathcal{N_{P_+P_-}}(\rho_3)$ (blue line) at stage 3  as a function of $p$. Realism cannot be maintained in Hardy's experiment (not even for $p=1$) and some aspects of realism-based nonlocality do manifest themselves (in pacific coexistence with local causality).}
\label{fig2}
\end{figure}

Application of the local-interaction model~\eqref{interaction} gives 
\be 
\ket{\Psi_3}=\ket{\phi_2^+}\ket{\phi_2^-}\ket{0}_\gamma+\tfrac{(1-\alpha)}{2}\ket{x}_+\ket{y}_-\ket{0}_\gamma-\tfrac{\beta}{2}\ket{0}_+\ket{0}_-\ket{2}_\gamma,
\label{Psi3}
\ee 
which is an entangled state. By tracing out the photon space, we are left with the state $\rho_3=\ket{\Theta_2}\bra{\Theta_2}+\tfrac{p}{4}\ket{0}\bra{0}$, where we have introduced, for notational simplicity, $\ket{\Theta_2}\equiv\ket{\phi_2^+}\ket{\phi_2^-}+\tfrac{(1-\alpha)}{2}\ket{x}_+\ket{y}_-$ and $\ket{0}\equiv\ket{0}_+\ket{0}_-$. The entanglement between radiation and matter can be estimated via the linear entropy $S_l$ of the reduced state, $E=S_l(\rho_3):=1-\wp(\rho_3)$, where $\wp(\rho_3):=\Tr(\rho_3^2)$ is the purity of the matter state. Via direct calculations we find $\wp(\rho_3)=\frac{1}{8}(8-4p+p^2)$ and $E=\frac{p}{2}\left(1-\frac{p}{4}\right)$, which are monotonic functions of $p$. Clearly, as entanglement increases, the purity of the matter state decreases. This is an important difference in relation to Hardy's approach: the photon constitutes a noisy channel for the positron-electron state. Given the symmetry of the system, irrealities are identical for the paths $\mathcal{P}_+$ and $\mathcal{P}_-$, the same applying for their local irrealities. After a lengthy and straightforward algebra, we arrive at
\beq 
\mathfrak{I}_\mathcal{P_\pm}(\rho_3)&=&-\ln{\sqrt{2}}+\tfrac{1}{4}\sum_{k=1}^2(-1)^k(2^k-p)\ln{(2^k-p)}, \\
\mathfrak{I}_\mathcal{P_\pm}(\rho_3^\pm)&=&-\tfrac{(6-p)}{4}\ln{2}-\tfrac{1}{16}\sum_{k=1}^2\sum_{j=1}^k\left[3(-1)^{k+1}+1\right]f_{jk}(p)\ln{\left[f_{jk}(p)\right]},
\eeq
where 
\be
f_{jk}(p):=2^k-p+(-1)^j\left[\frac{1+(-1)^k}{2}\right]\sqrt{8\left(1+\sqrt{1-p}\right)+p(p-4)}.
\ee 
For the contextual realism-based nonlocality we obtain
\be 
\mathcal{N_{P_+P_-}}(\rho_3)=-\ln{2}+\tfrac{1}{8}\sum_{k=0}^2\left[3(-1)^k-1\right](2^k-p)\ln{(2^k-p)}.
\ee 
The above quantities are graphed in Fig.~\ref{fig2} as a function of the annihilation probability $p$. Referring back to the fact that $p=0$ represents a setting in which both interferometers are isolated from each other, it is unsurprising that the irrealities corresponding to this regime recover the values obtained at stage 2 (both equal to $\ln 2$). For the other far end of the graph ($p=1$), which corresponds to Hardy's original picture, both irrealities remain positive, that is, there continues to be no element of reality associated with the paths [see Fig.~\ref{fig2}(a)]. The decrease in the irrealities with $p$ can be explained by the aforementioned decrease in the matter state purity, which is a direct consequence of the entanglement between matter and radiation. Fig.~\ref{fig2}(b) presents the contextual realism-based nonlocality at stage 3 as a function of $p$. The curve shows that the larger the probability of annihilation, the more sensitive the irreality of one particle to local measurements performed in the other. We recall that $\mathcal{N_{P_+P_-}}$ is not associated with nonlocal causality.

\subsection*{\it {\bf Stage 4.} Region after the final beam-splitters and before the detectors.}
\begin{figure}[htb]
\centering
\includegraphics[scale=0.45]{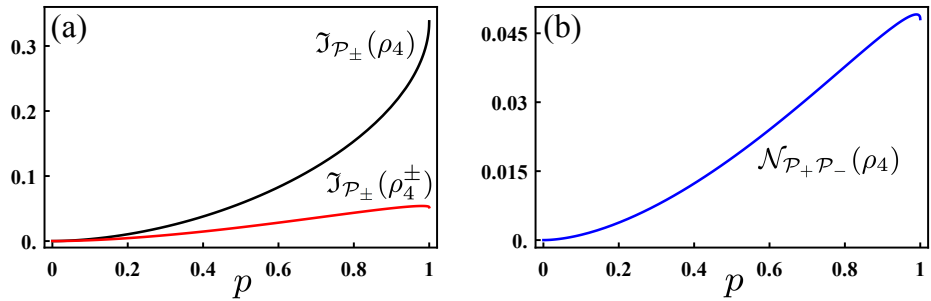}
\caption{\small (a)~Path irrealities $\mathfrak{I}_\mathcal{P_\pm}(\rho_4)$ (upper black line) and path local irrealities $\mathfrak{I}_\mathcal{P_\pm}(\rho_4^\pm)$ (lower red line) at stage 4 of Hardy's experiment as a function of the annihilation probability $p$. When $p=0$, the calibration of the interferometers ensures, via path interference, well-defined outcomes for the particles (no irreality). As $p$ increases, the interaction leads this balance to be disturbed and the particles to get correlated, which make the path irrealities $\mathfrak{I}_\mathcal{P_\pm}(\rho_4)$ increase. (b)~Contextual realism-based nonlocality $\mathcal{N_{P_+P_-}}(\rho_4)$ (blue line) at stage 3  as a function of $p$. Realism and locality appear only for $p=0$, which corresponds to the scenario with nonoverlapping interferometers. The abrupt bending in both $\mathfrak{I}_\mathcal{P_\pm}(\rho_4^\pm)$ and $\mathcal{N_{P_+P_-}}(\rho_4)$ in the domain close to $p=1$ derives from a subtle combination of behaviors of the entropic parcels underlying these quantities.}
\label{fig3} 
\end{figure}

Right after the particles are redirected by the mirrors, nothing happens from the viewpoint of irreality or nonlocality, since neither coherence nor correlations are altered. However, as the particles cross the latter set of beam-splitters, quantum coherence is inserted and the stage-3 scenario changes. To obtain the state in this region we apply the maps \eqref{mirror} and \eqref{b-s}, in this order, to $\ket{\Psi_3}$. The first term in Eq.~\eqref{Psi3} changes according to the maps $\ket{\phi_2^+}\mapsto -\ket{y}_+$ and $\ket{\phi_2^-}\mapsto-\ket{x}_-$, which correctly describe the occasion where the interferometers do not overlap. The last term in Eq.~\eqref{Psi3} does not change. After the system interacts with the mirrors, the intermediary term in Eq.~\eqref{Psi3} becomes $-\ket{y}_+\ket{x}_-\ket{0}_\gamma$, which is equal to $\ket{\Psi_1}$. Hence, after the passage through the final beam-splitters, this term evolves as $\ket{\Psi_1}\mapsto\ket{\Psi_2}=\ket{\phi_2^+}\ket{\phi_2^-}\ket{0}_\gamma$. We then obtain 
\be 
\ket{\Psi_4}=\ket{y}_+\ket{x}_-\ket{0}_\gamma-\tfrac{(1-\alpha)}{2}\ket{\phi_2^+}\ket{\phi_2^-}\ket{0}_\gamma-\frac{\beta}{2}\ket{0}_+\ket{0}_-\ket{2}_\gamma.
\ee 
Tracing out the photon space gives $\rho_4=\ket{\Theta_4}\bra{\Theta_4}+\tfrac{p}{4}\ket{0}\bra{0}$, where $\ket{0}\equiv\ket{0}_+\ket{0}_-$ and $\ket{\Theta_4}\equiv \ket{y}_+\ket{x}_--\tfrac{(1-\alpha)}{2}\ket{\phi_2^+}\ket{\phi_2^-}$. The purity of the positron-electron state and its entanglement with the photon space does not change with respect to the previous stage because there is no further interaction between these systems. As for the previous stage, here we succeeded to find analytical expressions for the irrealities and the contextual realism-based nonlocality, but now the resulting expressions are much more complicated and, for this reason, are omitted. These quantities are plotted in Fig.~\ref{fig3} as a function of the annihilation probability $p$. In the vicinity of $p=0$ one has, for the leading order in $p$,
\beq 
\mathfrak{I}_\mathcal{P_\pm}(\rho_4)&\cong& \Big(1+\ln{32}-\ln{p^2}\Big)\frac{p^2}{32}, \\
\mathfrak{I}_\mathcal{P_\pm}(\rho_4^\pm)&\cong& \Big(1+\ln{16}-\ln{p^2}\Big)\frac{p^2}{64}, \\
\mathcal{N_{P_+P_-}}(\rho_4)&\cong& \Big(1+\ln{4}-\ln{p^2}\Big)\frac{p^2}{64}.
\eeq 
These results reveal the (smooth) form through which the irrealities and the contextual realism-based nonlocality vanish with $p$. Moreover, along with Fig.~\ref{fig3}, these formulas show that realism and locality strictly emerge only if $p=0$, an instance that is equivalent to having no overlap between the interferometers. Again, it is noteworthy the fact that even though $\mathcal{N_{P_+P_-}}(\rho_4)>0$ for $p>0$, no action at a distance is assumed whatsoever.

It is interesting to note that the probability of getting simultaneous clicks in the dark detectors, X$_+$ and Y$_-$, is given by 
\be 
p_\text{dark}=\left|\bra{\Psi_4}\left(\ket{x}_+\ket{y}_-\ket{0}_\gamma\right)\right|^2=\frac{\left(1-\sqrt{1-p} \right)^2}{16}.
\ee 
This monotonically increasing function of $p$ turns out to describe as well the probabilities of simultaneous clicks in both X$_+$ (dark detector) and X$_-$, and  Y$_-$ (dark detector) and Y$_+$. Therefore, the probability of at least one dark detector clicking is $3\,p_\text{dark}$. This predicts that the paradox emerges for all $p>0$, being maximally accentuated in Hardy's original setting ($p=1$). Most importantly, it comes from the above results that whenever the paradox emerges ($p_\text{dark}>0)$, realism becomes indefensible anywhere else but at the first stage. 

\subsection{Discussion}

The usual interpretation of Hardy's experiment is fundamentally {\it realistic} and, as a consequence, willing to accept {\it nonlocal causality} (actions at a distance). When the dark detectors click, the absence of annihilation is claimed to imply that the particles must have taken realistic distinct paths and interacted at a distance, otherwise these detectors should have remained dark. In this model, {\it retrodictions} (predictions about the past) are necessary. In fact, upon clicks, inferences are made about the past trajectories, which are assumed to come into existence even one knowing that the particles traveled the interferometers in quantum superpositions. The scenario implied by this interpretation is presented in Table~\ref{table1}(a). Because the trajectories of the particles are merely ``revealed'' by the final measurements, realism is satisfied at every stage of the experiment. Then, at stage 3, when the particles presumably interact even being far apart from each other, the local causality hypothesis is violated and Bell's nonlocality is demonstrated without inequalities. 

In the previous section, by use of a recently introduced framework that allows for the quantification of reality, we have seen that Hardy's experiment can be analyzed through a conceptually different narrative, here named ``irrealistic'' model [see Table\ref{table1}(b)]. Let us confine our discussion to Hardy's setting, $p=1$. Already at stage 2, coherence (quantum superposition) is introduced by the beam-splitters in  particles' paths. According to the irreality measure \eqref{frakI}, local irreality suffices to establish a negation of the realism hypothesis [Eq.~\eqref{BAcriterion}]. In fact, at this stage the paths irrealities are maximal. When local interaction takes place, at stage 3, quantum correlations are generated between the particles (also with radiation). On the one hand, these correlations reduce local irreality, because coherence is destroyed in the subsystems; on the other hand, they contribute positively to irreality. The net effect is such that paths' irrealities are no longer maximal but are still greater than zero, thus ruling out realism.

Another interesting aspect of the irrealistic model is the persistence of nonlocal aspects. Even though only local interactions are assumed throughout the experiment---which explains the check-marks for local causality in Table~\eqref{table1}(b)---realism-based nonlocality does manifest itself at stages 3 and 4. Within the operational scheme devised by Bilobran and Angelo~\cite{bil}, this means that changes in the positron path-irreality due to unrevealed measurements performed on the electron can be experimentally detected. For the Hardy experiment, the conceptual picture can be as follows. Irreality for the positron path can be thought of as meaning that the particle actually travels {\it both} arms simultaneously, like a wave. In this capacity, the positron can {\it locally} interact with the electron, which also occupies both arms. A distinctive advantage of the present model is that, being able to quantitatively assess the irrealism and the realism-based nonlocality associated with the quantum state at every instant of time, it allows for self-consistent causal analyses along the whole time evolution of the system, thus avoiding retrodictions.

\begin{table}[htb]
\centering
(a) Realistic Model
\vskip1mm
\begin{tabular}{|c|c|c|c|c|}
\hline\hline
              & Stage 1 & Stage 2 & Stage 3 & Stage 4 \\ \hline\hline
Realism & $\checkmark$ & $\checkmark$ & $\checkmark$ & $\checkmark$ \\ 
Local Causality& $\checkmark$ & $\checkmark$ & $\times$ & $\checkmark$ \\
\hline\hline
\end{tabular}
\vskip5mm
(b) ``Irrealistic'' Model
\vskip1mm
\begin{tabular}{|c|c|c|c|c|}
\hline\hline
              & Stage 1 & Stage 2 & Stage 3 & Stage 4 \\ \hline\hline
Realism & $\checkmark$ & $\times$ & $\times$ & $\times$ \\ 
Local Causality& $\checkmark$ & $\checkmark$ & $\checkmark$ & $\checkmark$ \\
\hline\hline
\end{tabular}
\vskip5mm
\caption{(a)~In the realistic interpretation of Hardy's experiment ($p=1$), only the hypothesis of local causality is violated. When the dark detectors click, retrodiction is used to assume that the particles have interacted at a distance while traveling through spatially separated realistic paths at stage 3. (b)~In the ``irrealistic'' model, wherein local causality is assumed, the dynamical development of quantum superpositions and correlations implies the negation of realism already at stage 2.}
\label{table1}
\end{table}

\section{Conclusion}
\label{conc}

Given the fact that quantum mechanics has never failed in its predictions about nature, one may fairly take the stance that the theory is correct, in which case there should be no paradox in its predictions, even if they may prove counterintuitive. Any conflicts can then be attributed to tacit attempts to understand nature in terms of nonquantum principles. That is exactly what happens in Hardy's experiment when we tacitly use retrodiction to presume that the particles transited in a localized manner through the arms of the interferometers. Actually, right from the start, the assumption of realistic trajectories stands at odd with the interpretation we commonly attach to superposition states, such as those generated by the first set of beam-splitters. The irrealistic model introduced here quantitatively demonstrates the point: the typical scenario inside the interferometers is of clear violation of realism. This conclusion is corroborated by the realism-based nonlocality, whose manifestation is conditioned to quantum irrealism. Abandoning realism, one  admits particle delocalization and, with that, explains the disturbance between the particles without invoking nonlocal causality. Then, accepting that this is the natural state of affairs, the paradox disappears.

Putting in perspective, this work does not disprove the usual realistic interpretation, but proposes an alternative to it. The common view, advocated by Hardy and consonant with Bell's theorem, consists of keeping realism and abandoning local causality, with basis on retrodiction. The alternative one, discussed in this work and suggested by Hardy himself, keeps local causality and abandons realism. Nevertheless, a residual nonlocality (different from Bell's nonlocality) emerges which is intimately related to irrealism: because the particles may be at various places simultaneously, they may interact locally at various places simultaneously, which can result in irreality alterations induced by remote measurements. Even though the state of the art experiments are not able to rule out one of the above models, it seems fair to conclude that the dissolution of Hardy's paradox irremediably demands the abolishment of at least one of our deeply-rooted preconceptions about nature.

\begin{acknowledgements}
The authors acknowledge CNPq (Brazil) and the National Institute for Science and Technology of Quantum Information (CNPq, INCT-IQ 465469/2014-0).
\end{acknowledgements}



\end{document}